\documentclass[12pt]{article}
\usepackage{amsmath,amssymb,amsfonts,bm,marginnote,cite,graphicx,
,slashed,color}
\usepackage[colorlinks=true, linkcolor=blue, citecolor=blue, linktoc=all]{hyperref}
\usepackage[usenames,dvipsnames]{xcolor}

\edef\marginnotetextwidth{\the\textwidth}

\newcommand{\thistitle}{
Intrinsic non-commutativity of closed string theory}
\newcommand{\addresspi}{
	Perimeter Institute for Theoretical Physics, 
	31 Caroline St. N.,  Waterloo ON, N2L 2Y5, Canada
	}
\newcommand{\addressuiuc}{
	Department of Physics, University of Illinois,
 	1110 West Green St., Urbana IL 61801, U.S.A.
	}
\newcommand{\addressvt}{
	Department of Physics, Virginia Tech,  
	Blacksburg VA 24061, U.S.A.
	}
\newcommand{\emaillf}{lfreidel@perimeterinstitute.ca}
\newcommand{\emailrgl}{rgleigh@illinois.edu}
\newcommand{\emaildm}{dminic@vt.edu}

\newcommand{\be}{\begin{equation}}
\newcommand{\ee}{\end{equation}}
\newcommand{\beq}{\begin{eqnarray}}
\newcommand{\eeq}{\end{eqnarray}}
\newcommand{\bea}{\begin{eqnarray}}
\newcommand{\eea}{\end{eqnarray}}
\newcommand{\beqn}{\begin{eqnarray}}
\newcommand{\eeqn}{\end{eqnarray}}

\newcommand{\X}{\mathbb{X}}

\newcommand{\Z}{\mathbb{Z}}

\newcommand{\K}{\mathbb{K}}

\newcommand{\Pm}{\mathbb{P}}

\def\pa{\partial}

\newcommand{\rd}{\mathrm{d}}
\def\dd{\!\cdot \!}

\def\dd{\!\cdot \!}

\def\s{\sigma}

\def\dd{\!\cdot \!}

\def\s{\sigma}

\def\tx{\tilde{x}}

\def\tk{\tilde{k}}

\def\tq{\tilde{q}}
\def\tp{\tilde{p}}

\renewcommand{\thefootnote}{\fnsymbol{footnote}}
\begin{document}

\title{\thistitle}
\author{
	{Laurent Freidel$^{a}\footnote{\emaillf}$, Robert G. Leigh$^{b}\footnote{\emailrgl}$\ and Djordje Minic$^{c}$\footnote{\emaildm}}\\
	\\
	{\small ${}^a$\emph{\addresspi}}\\ 
	{\small ${}^b$\emph{\addressuiuc}}\\ 
	{\small ${}^c$\emph{\addressvt}}\\
\\}
\maketitle\thispagestyle{empty}
\vspace{-5ex}
\begin{abstract}
We show that the proper interpretation of the cocycle operators appearing in the physical vertex operators of compactified strings is that the closed string target is non-commutative. We track down the appearance of this non-commutativity to the Polyakov action of the flat closed string in the presence of translational monodromies (i.e., windings). In view of the unexpected nature of this result, we present detailed calculations from a variety of points of view, including a careful understanding of the consequences of mutual locality in the vertex operator algebra, as well as a detailed analysis of the symplectic structure of the Polyakov string. 
We also underscore why this non-commutativity was not emphasized previously in the existing literature.
This non-commutativity can be thought of as a central extension of the zero-mode operator algebra, an effect set by the string length scale -- it is present even in trivial backgrounds. Clearly, this result indicates that the $\alpha'\to 0$ limit is more subtle than usually assumed.
\end{abstract}
\bigskip

\setcounter{footnote}{0}
\renewcommand{\thefootnote}{\arabic{footnote}}

\section{Introduction}

One of the annoying technicalities of string theory is the presence of co-cycles in the physical vertex operators. In the standard account, these co-cycles are required in order to maintain locality on the worldsheet, i.e., to obtain mutual locality of physical vertex insertions. For example, they appear in standard discussions \cite{Polchinski:1998rq} of compactified strings, and rapidly lead to both technical and conceptual issues.  
In this paper, we re-analyze these issues carefully, and show that the space of string zero modes surprisingly is best interpreted as non-commutative, with the scale of non-commutativity set by $\alpha'$. A by-product of this realization is that the operator algebra becomes straightforward (albeit with a non-commutative product), with no need for co-cycles. This is not inconsistent with our usual notion of space-time in decompactification limits, but it does significantly impact the interpretation of compactifications in terms of local effective field theories. This is a central ingredient that has been overlooked in any of the attempts at duality symmetric formulations of string theory. Indeed, in a follow-up paper we will show that one can obtain a simple understanding of exotic backgrounds such as asymmetric orbifolds \cite{Narain:1986qm} and T-folds \cite{Hull:2006va}. 

Much of the usual space-time interpretation that we use in string theory is built in from the beginning. Its origins, for example, as an S-matrix theory in Minkowski space-time is emblematic of its interpretation in terms of a collection of particle states propagating in a fixed space-time background. We typically view other solutions of string theory in a similar way, with a well-defined distinction between what is big and what is small. Each such case can be viewed as a classical or semi-classical approximation to a deeper quantum theory in which the notion of a given space-time is {\it not} built in from the beginning, but is an emergent property of a given classical limit. It is natural to ask under what circumstances a local effective field theory is obtained. Of course, we know many such instances, and we also know many examples where this does not occur, such as cases where non-commutative field theories are thought to emerge. Perhaps the avatar for the absence of a fixed space-time picture is given by duality-symmetric formulations (of which double field theories \cite{Hull:2009mi} and our own 
metastring theory \cite{Freidel:2013zga, Freidel:2014qna, Freidel:2015pka, Freidel:2015uug, Freidel:2016pls, Freidel:2017xsi}, are examples). We are in fact working towards a new notion of {\it quantum space-time}, in which non-commutativity plays a central role, much as it does in ordinary quantum mechanics. In the present paper then, we uncover an important step towards such an understanding of quantum space-time.

\section{Classical Structure of Compact String}

To begin, we review various details of the free bosonic string. The majority of this material is standard fare, and we use this section to set-up our conventions, following closely the notation in Ref. \cite{Polchinski:1998rq}. The reader familiar with these details might wish to skip this overview and jump to equation (\ref{cocyclecon}).

What we are interested in is the canonical analysis of the string in which translational monodromy (winding) is allowed
\beq
X (\tau,\sigma + 2 \pi) = X(\tau,\sigma) + 2 \pi \alpha'\tp, \qquad \alpha'\tp\in\tilde \Gamma ,
\eeq
while paying particularly close attention to the zero modes.
In order to describe these boundary conditions, we have introduced a lattice\footnote{ A lattice is an abelian group equipped with a metric. This includes non-compact directions if we take the distance between two points of the lattice to zero. In our presentation we always think of non-compact directions as being decompactified.}   $\tilde\Gamma$. 
Of course 
 $\tilde{\Gamma}$ represents the lattice of winding modes due to the fact that that we are describing strings propagating on a compact geometry, $T=\mathbb{R}^d/\tilde\Gamma$. 
It is then often assumed that the structure of vertex operators and the effective description of the string in terms of field theory inherits the naive geometrical structure of the string and that the string compactification implies an effective Kaluza-Klein description involving the same  compact geometry. Our work will show that this traditional point of view can be misleading and that is not a generic message of string theory.

 We emphasize  that such an interpretation relies heavily on an assumed structure for the zero modes of the string, namely that they can be thought of as coordinates on $T$, along with their conjugate variables. This is of course precisely what is relevant in the non-compact case, the zero modes generating a Heisenberg algebra associated with the non-compact space-time. We are interested here in a careful analysis of the compact case, with generic periodic $\tilde\Gamma$. 
As is well known, when we turn on $\hbar$ another lattice appears, the lattice  of momenta $\Gamma$, dual to the winding lattice, $\Gamma=\tilde\Gamma^*$. Since our goal is to understand the geometry behind the presence of these two lattices it is interesting to note that one is quantized classically and the other one quantum mechanically and it will be useful to unify these two lattices into one denoted as $\Lambda \equiv \Gamma\oplus \tilde{\Gamma}$. The duality pairing between $\Gamma$ and $\tilde{\Gamma}$ equips $\Lambda$ with a canonical metric that will play a key role in our story.
  
Let us first review the usual classical solutions, again essentially to set notation. We work on a cylindrical worldsheet, which we think of as cut along $\sigma=0$ and unrolled,\footnote{Implicit here is the independence on where we cut open the cylinder. This will become an issue to be checked in the course of the construction of the symplectic structure, which we present in a separate section.} and the general classical solutions can then be written 
\beq
X(\tau,\sigma) =  X_R(\tau+\sigma)+X_L(\tau-\sigma) ,
\eeq
with\footnote{The Euclidean continuation, as in \cite{Polchinski:1998rq}, is that left-movers are holomorphic ($\tau-\sigma \to -i\ln z$) and right-movers anti-holomorphic, $\tau+\sigma\to -i\ln\bar z$.}
\beqn
X_L(\tau-\sigma) &=& x_L + \frac{\alpha'}{2} p_L (\tau-\sigma) +i \lambda \sum_{m= - \infty}^{\infty}  \frac{1}{m}\alpha_m e^{-im(\tau-\sigma)}
\\
X_R(\tau+\sigma) &=& x_R + \frac{\alpha'}{2} p_R (\tau+\sigma) +i \lambda \sum_{m= - \infty}^{\infty} \frac{1}{m}\tilde{\alpha}_me^{-im(\tau+\sigma)},
\eeqn
where we have introduced the string length scale
\be
\lambda\equiv \sqrt{\frac{\hbar\alpha'}{2}},
\ee 
with  $\alpha'$ the ``string slope'' which has dimensions of length over energy.
We see that 
the lattice vectors are given by $\tp=\frac{p_R-p_L}{2}$. Similarly, we define $p=\frac{p_R+p_L}{2}$.  The zero modes $(p,\tp)$ can be extracted via
\beq
p=\frac{1}{2\pi\alpha'}\int_{0}^{2\pi}d\sigma\ \pa_\tau X,\qquad \tp=\frac{1}{2\pi\alpha'}\int_{0}^{2\pi}d\sigma\ \pa_\sigma X.
\eeq
Thus, we have the familiar situation of the zero mode sector being coordinatized by $(x_L,p_L,x_R,p_R)$, all independent. We will be interested in the structure of physical (vertex) operators and their algebra, as well as the symplectic structure of the theory. Typically in the literature, it is taken as self-evident that $(x_L,p_L)$ and $(x_R,p_R)$ generate two commuting Heisenberg algebras when the theory is quantized. We will discuss this carefully in a separate section in a detailed analysis of the symplectic structure, but now we will need a somewhat weaker assumption that $(p_L,p_R)$ commute, which as we will see, is borne out by the symplectic analysis.

Given that $X_L(\tau-\sigma)$ and $X_R(\tau+\sigma)$ are independent of one another, 
it is natural to also introduce 
\be
\tilde{X}(\tau,\sigma)=X_R(\tau+\sigma)-X_L(\tau-\sigma),
\ee 
which is such that $\pa_\s \tilde{X}=\pa_\tau X$, 
and the corresponding classical solution may be written as\footnote{Here, we have defined
\beq
\hat Q_-(\tau-\sigma)=i \lambda \sum_{m= - \infty}^{\infty}  \frac{1}{m}\hat \alpha_m e^{-im(\tau-\sigma)},\qquad
\hat Q_+(\tau+\sigma)=i \lambda \sum_{m= - \infty}^{\infty}  \frac{1}{m}\hat{\tilde{\alpha}}_me^{-im(\tau+\sigma)}.
\eeq}
\beqn \label{mode1}
X(\tau,\sigma) &=& x +\alpha' (p\tau+\tp\sigma) 
 +Q_-(\tau-\sigma)+Q_+(\tau+\sigma)
\\
\tilde{X}(\tau,\sigma) &=& \tx +\alpha' (\tp\tau+p\sigma) 
-Q_-(\tau-\sigma)+Q_+(\tau+\sigma) \label{mode2},
\eeqn
where $x=x_R+x_L$ and $\tx=x_R-x_L$. We regard $X(\tau,\sigma), \tilde{X}(\tau,\sigma)$ as a reorganization of $X_L(\tau-\sigma), X_R(\tau+\sigma)$. In particular, the zero modes $(x,\tx,p,\tp)$ should be considered independent, being linear combinations of $(x_L,p_L,x_R,p_R)$. 

The field $\tilde X(\tau,\sigma)$ has a well-known significance in the limit when the radius of compactification is sent to zero. In this limit, the $(x,p)$ zero modes decouple, and $\tilde X(\tau,\sigma)$ is the field that the string couples to, with $\tx$ coordinatizing the (T-dual) non-compact geometry. The momentum $p$ plays the role of the translational monodromy of $\tilde X(\tau,\sigma)$, i.e.,
\beq
\tilde X (\tau,\sigma + 2 \pi) = \tilde X(\tau,\sigma) + 2 \pi\alpha' p.
\eeq
It should be clear then that what we mean by the space-time that the string ``sees'' should not be taken to be given by $X(\tau,\sigma)$. The meaning of $X(\tau,\sigma)$ will change in different limits. This is the essence of T-duality. The field $X(\tau,\sigma)$ has a space-time interpretation in the original Polyakov path integral only in the limit of large radius.

Furthermore, we note that 
\beq\label{Xduality}
dX=*d\tilde{X},\qquad i.e. \quad \pa_\tau X=\pa_\s\tilde{X},\quad \pa_\sigma X=\pa_\tau\tilde{X},
\eeq 
and so T-duality, which interchanges $X$ and $\tilde X$, is associated with Hodge duality on the worldsheet, as far as the currents are concerned. The momentum field conjugate to $X(\tau,\sigma)$ is of course the momentum density $P(\tau,\sigma)=\frac{1}{2\pi\alpha'}\pa_\tau X(\tau,\sigma)$. Similarly, it is convenient to introduce the winding density $\tilde{P}(\tau,\sigma)=\frac{1}{2\pi\alpha'}\pa_\tau \tilde{X}(\tau,\sigma)$.

Our intention in this discussion is to highlight the zero modes. Of course, it is often assumed that the zero mode $x$ plays a preferred role, being interpreted as a coordinate in the target space, while $\tx$ is immaterial as it does not appear in the action. 
 In fact, neither $x$ nor $\tx$ appear in the abelian currents $dX$ and the string action density. However, they do enter the theory in the vertex operator algebra, and so we are led to study them more carefully. In the present context {\it both} $(x_L,x_R)$ (equivalently $(x,\tx)$) do appear in vertex operators, and we will take pains to treat them with care. It is implicit in this analysis that $(x_L, x_R)$  (equivalently $(x, \tx)$) are independent; it is only in the $R\to\infty$ ($R\to 0$) limit that a projection $\tp\to 0$ ($p\to 0$) on the spectrum is induced, at which point $\tx$ ($x$) decouples. 

\section{The Vertex Algebra and Mutual Locality}

As we mentioned earlier, it is often assumed \cite{Polchinski:1998rq} from the outset that the zero modes satisfy Heisenberg algebras
\beqn
[\hat{x}, \hat{p}]= i \hbar = [\hat{\tx},\hat{\tp}], \label{HeisAlgs}
\eeqn
(equivalently, $[\hat{x}_L,\hat{p}_L]=i\hbar=[\hat{x}_R,\hat{p}_R]$.)  We will argue below that there is an important subtlety here, related to the commutativity of $\hat{x}_L, \hat{x}_R$, that has been overlooked. To get there, we begin by considering local vertex operators and the requirement of mutual locality. It is convenient to revert to Euclidean methods, rewriting\footnote{We will see that the oscillators satisfy the usual commutation relations,
\beq
[\hat\alpha_n^a,\hat\alpha_m^b ]=   n h^{ab} \delta_{n+m}, \qquad \hat\alpha^\dagger_n = \hat\alpha_{-n},
\eeq
and similarly for $\tilde{\alpha}$, where $h$ is the space-time Lorentz metric.
}
\beqn
\hat X_L(z) = \hat x_L(z) +i \lambda \sum_{m= - \infty}^{\infty}{\kern-9pt}{}^{'} \frac{1}{m }\hat{\alpha}_mz^{-m}\label{leftholo}
,\qquad
\hat X_R(\bar{z}) = \hat x_R(\bar z) +i \lambda \sum_{m= - \infty}^{\infty}{\kern-9pt}{}^{'} \frac{1}{m}\hat {\tilde{\alpha}}_m \bar{z}^{-m},
\eeqn
where the prime in the sum means that we omit $m=0$. Also we introduced the zero modes functionals
\be  
\hat x_L(z):=\hat x_L-i\frac{\alpha'}{2}\hat p_L\ln z,\qquad
\hat x_R(\bar{z}):=\hat x_R-i\frac{\alpha'}{2}\hat p_R\ln \bar{z}.
\ee 

Now we are interested in physical vertex operators that have well-defined conformal properties. For simplicity, we concentrate on the tachyon operators\footnote{As is standard, $::$ refers to the usual oscillator normal ordering. This removes, for example, overall factors in eq. (\ref{splitosc}).}
\be
\hat{W}_{\K}(z,\bar z)=  : 
e^{ik_L\cdot \hat X_L(z)+ik_R\cdot \hat X_R(\bar z)}: \hat{C}_{\K},
\ee
where hat is used to denote operators, $\K$ stands for the pair $(k_L,k_R)$ (or equivalently $(\tk,k)$) and the double dots denote a normal ordering prescription that extends to the zero modes. Implicit in this is the assumption that $\hat{\tp}$ and $\hat{p}$ commute, and thus we choose to label operators by their eigenvalues $\hbar (\tk,k)$, where the wavelength vectors $k$ and $\tk$  have units of 1/length.

 As noted in \cite{Polchinski:1998rq}, we must take care of the branch cuts in the logarithms appearing in (\ref{leftholo}), and in particular, of their effect on the commutation properties of the vertex operators, since $\hat{X}_L$ and $\hat{ X}_R$ are not themselves well-defined operators. Correspondingly, we have included an operatorial co-cycle factor $\hat{C}_{\K}$ in the vertex operators. This factor is independent of $(z,\bar z)$ and is assumed to be a function of $(\hat p_L,\hat p_R)$, in other words $\hat{C}_{\K}={C}_{\K}(\hat p_L,\hat p_R)$. As we are going to see, mutual locality drastically  constrains the form of this operator.
 
The short-distance singularities at finite separation are of the form
\be
X_L(z_1) X_L(z_2) \sim - \lambda^2 \ln{z_{12}}, \quad X_R(\bar{z}_1) X_R(\bar{z}_2) \sim - \lambda^2 \ln{\bar{z}_{12}}, \quad X_L(z_1) X_R(\bar{z}_2) \sim 0.
\ee
We will separate the zero-modes from the oscillator parts, and so we write
\beqn
\hat{W}_{\K}(z,\bar z)&=&
 \hat U_{\K}(z,\bar z) \hat{V}_{\K}(z,\bar z)
 = \hat U_{\K}(z,\bar z) \hat{V}_{k_L}(z)\hat{V}_{k_R}(\bar{z}),
\eeqn
where $U_{\K}(z,\bar z)$ contains the zero-modes, and $V_{\K}$ the normal ordered oscillators
\beqn
\hat{V}_{k_L}(z) = 
\hat{E}_{-k_L}^\dagger(z)\, \hat{E}_{k_L}(z),\qquad
\hat{V}_{k_R}(\bar{z}) = 
\hat{\bar{E}}_{-k_R}^\dagger(\bar{z})\,  \hat{\bar{E}}_{k_R}(\bar{z}).
\label{splitosc}
\eeqn
Here we have introduced\footnote{ The dagger operation is the BPZ conjugation $[\Phi(z,\bar{z})]^\dagger = \Phi^\dagger(z^{-1},\bar{z}^{-1})$, which is consistent with the Lorentzian definition $(z,\bar{z})\to (e^{i(\tau-\s)}, e^{i(\tau+\s)})$ and $\alpha_n^\dagger=\alpha_{-n}$. The definition implies that $V_{k_L}$ are unitary operators.}
\be
\hat{E}_{k_L}(z):=e^{-\lambda\sum_{m=1}^\infty \frac{1}{m} k_L\cdot \hat\alpha_{m}z^{-m}},\qquad \hat{E}^\dagger_{-k_L}(z):=e^{\lambda\sum_{m=1}^\infty \frac{1}{m} k_L\cdot \hat\alpha_{-m}z^m},
\ee
and, similarly for the  right-movers. We also define the zero mode operator 
\beq\hat U_{\K}(z,\bar z):=\hat U_{\K}\,z^{\frac{\alpha'}{2} k_L\cdot\hat{p}_L}\, \bar{z}^{\frac{\alpha'}{2} k_R\cdot\hat{p}_R} .
\eeq

 We will discuss its detailed form later, but for now $\hat U_{\K}$ can be considered as an exponential operator depending 
 on the zero modes $\hat x_{L,R}$
 \be\label{Udef}
 \hat U_{\K} =  e^{ik_L\cdot \hat x_L+ik_R\cdot \hat x_R}\,{C}_{\K}(\hat{p}_L,\hat{p}_R).
 \ee
 $\hat U_{\K}$ needs to satisfy two properties. First, the product of two such operators needs to close. We therefore suppose that products may involve a co-cycle factor $\epsilon_{\K,\K'}\in \mathbb{C}^*$
\beq\label{defcocycle}
\hat U_\K \hat U_{\K'}=\epsilon_{\K,\K'} \hat U_{\K+\K'},
\eeq
and the associativity of the product implies that the cocycles must satisfy 
\beq\label{cocyc}
\epsilon_{\mathbb{K},\mathbb{L}}
\epsilon_{\mathbb{K}+\mathbb{L},\mathbb{M}}=
\epsilon_{\mathbb{L},\mathbb{M}}
\epsilon_{\mathbb{K},\mathbb{L}+\mathbb{M}}.\label{cocycle}
\eeq
Second, $\hat U_{\K}$ will have non-trivial commutation with functions of $\hat p_{L,R}$ and in particular we will need
\beqn\label{reorderingJunk}
z^{\frac{\alpha'}{2} k_L\cdot\hat{p}_L}\bar{z}^{\frac{\alpha'}{2} k_R\cdot\hat{p}_R} \hat{U}_{\K'}
&=&z^{\lambda^2k_L\cdot k'_L}\bar{z}^{\lambda^2k_R\cdot k'_R}\,
\hat{U}_{\K'}\, z^{\frac{\alpha'}{2} k_L\cdot\hat{p}_L}\bar{z}^{\frac{\alpha'}{2} k_R\cdot\hat{p}_R},
\eeqn
which follows from eq. (\ref{HeisAlgs}).
Now we compute
\beqn
\hat{W}_{\K}(z_1,\bar z_1)\hat{W}_{\K'}(z_2,\bar z_2)
&=&
\hat{U}_{\K} z_1^{\frac{\alpha'}{2} k_L\cdot\hat{p}_L}\bar{z}_1^{\frac{\alpha'}{2} k_R\cdot\hat{p}_R}
\hat{U}_{\K'} z_2^{\frac{\alpha'}{2} k'_L\cdot\hat{p}_L}
\bar{z}_2^{\frac{\alpha'}{2} k'_R\cdot\hat{p}_R}
\\&&\times
\hat{E}_{-k_L}^\dagger(z_1)\, \hat{E}_{k_L}(z_1)\hat{E}_{-k'_L}^\dagger(z_2)\, \hat{E}_{k'_L}(z_2)
\hat{E}_{-k_R}^\dagger(\bar z_1)\, \hat{E}_{k_R}(\bar z_1)\hat{E}_{-k'_R}^\dagger(\bar z_2)\, \hat{E}_{k'_R}(\bar z_2). \nonumber
\eeqn
We will perform re-orderings on the right-hand side so that it can be written in terms of expressions involving $\K+\K'$. We note that  
\beqn
\hat{E}_{k_L}(z_1)\hat{E}^\dagger_{-k'_L}(z_2)&=&
e^{-\lambda^2k_L\cdot k'_L\sum_{n=1}^\infty\frac{1}{n} \left(\frac{z_2}{z_1}\right)^n}
\hat{E}^\dagger_{-k'_L}(z_2)\hat{E}_{k_L}(z_1)\label{ordersum}\\
&=&
\left(1-\frac{z_2}{z_1}\right)^{\lambda^2k_L\cdot k'_L}
\hat{E}^\dagger_{-k'_L}(z_2)\hat{E}_{k_L}(z_1)\label{rearrangeL},
\eeqn
the sum converging uniformly within the disc $|z_2|<|z_1|$. 
Putting everything together, we find
\beq\label{OPE1}
\hat{W}_{\K}(z_1,\bar{z}_1)\hat{W}_{\K'}(z_2,\bar{z}_2) =
z_{12}^{ \lambda^{2} k_L\cdot k'_L}
\bar{z}_{12}^{ \lambda^{2}  k_R\cdot k'_R} \epsilon_{\K,\K'}\, 
\hat{U}_{\mathbb{K}+\mathbb{K}'}: \hat{V}_{\K} (z_1,\bar{z}_1) \hat{V}_{\K'}(z_2,\bar{z}_2) : ,
\eeq
where $z_{12}=z_1-z_2$. 
Again, this is valid for the time-ordering $|z_2|<|z_1|$. For the exterior of the disk, $|z_1|<|z_2|$, the sums converge for the opposite operator ordering
\beq\label{OPE2}
\hat{W}_{\K'}(z_2,\bar{z}_2)\hat{W}_{\K}(z_1,\bar{z}_1) =
z_{21}^{ \lambda^{2} k_L\cdot k'_L}
\bar{z}_{21}^{ \lambda^{2}  k_R\cdot k'_R} \epsilon_{\K',\K}\, 
\hat{U}_{\mathbb{K}+\mathbb{K}'}: \hat{V}_{\K} (z_1,\bar{z}_1) \hat{V}_{\K'}(z_2,\bar{z}_2) : .
\eeq
Equations (\ref{OPE1}, \ref{OPE2}) lead to the radial ordered correlation function and also to the operator product expansion. Indeed, the normal-ordered product may be expanded to a series of local operators, weighted by integer powers of $z_{12}$ and $\bar{z}_{12}$, and thus a given term in the series will scale (in the case of (\ref{OPE1})) as $z_{12}^{ \lambda^{2} k_L\cdot k'_L+m}\bar{z}_{12}^{ \lambda^{2} k_R\cdot k'_R+n}$, for $m,n\in\Z$. 
 
Mutual locality is the requirement that operators commute at space-like separation. We can investigate this by comparing the limits of (\ref{OPE1}) and (\ref{OPE2}) as $|z_1|\to |z_2|$ (for some spatial separation $\sigma_{12}=\sigma_1-\sigma_2$). It is most straightforward to go back and write the expression in terms of the summation appearing in (\ref{ordersum}). Then one finds in the limit
\beq
\hat{W}_{\K}(|z|,\sigma_1)\hat{W}_{\K'}(|z|,\sigma_2)
=
 \frac{\epsilon_{\K,\K'}}{ \epsilon_{\K',\K}} \left(e^{i\sigma_{12}+\sum_{n\neq 0}\frac{1}{n}e^{in\sigma_{12}}}\right)^{\lambda^{2} (k_L\cdot k'_L-k_R\cdot k'_R)}
\hat{W}_{\K'}(|z|,\sigma_2)\hat{W}_{\K}(|z|,\sigma_1).
\eeq
We write this result in the more compact form
\beqn\label{orderops}
\hat{W}_{\K}(|z|,\sigma_1)\hat{W}_{\K'}(|z|,\sigma_2)=
\frac{\epsilon_{\K,\K'}}{ \epsilon_{\K',\K}}\ e^{-i2 \eta(\lambda \K,\lambda \K')\theta(\sigma_{12})}\
\hat{W}_{\K'}(|z|,\sigma_2)\hat{W}_{\K}(|z|,\sigma_1),
\eeqn
where we recognize $\theta(\sigma)$ as the staircase distribution
\beq
\theta(\sigma)=\sigma-i\sum_{n\neq 0}\frac{e^{in\sigma}}{n}.
\eeq
This distribution satisfies three key properties: its derivative is proportional to the Dirac distribution,
 $\theta'(\sigma)=2\pi\delta(\sigma)$, it is quasi-periodic with period $2\pi$,  $\theta(\sigma+2\pi n)=\theta(\sigma)+2\pi n$, and it is odd, $\theta(-\sigma)=-\theta(\sigma)$.  
 In particular this implies that  $\theta(\sigma)$ is valued in 
 $\pi \mathbb{Z}$ for $\sigma \in \mathbb{R}\backslash \mathbb{Z}$.
We have also introduced a bilinear form $\eta$ in order to write the result in a compact form\footnote{$k=\frac12(k_R+k_L)$ and $\tk=\frac12(k_R-k_L)$ have units of $1/$length. Explicitly
\beq
\K^A=\begin{pmatrix}\tilde k^a\cr k_a\end{pmatrix},\qquad
\eta_{AB}= \begin{pmatrix} 0&{\bf 1}\cr {\bf 1}&0\end{pmatrix},
\eeq
and $\lambda \mathbb{K}$ form a lattice, 
where $\lambda^2=\hbar\alpha'/2$. 
}
\beq\label{opeloc1}
\eta(\K,\K')
:=\frac12(k_R\cdot k'_R-k_L\cdot k'_L)
= (k\cdot\tk'+\tk\cdot k').
\eeq
We can now infer two important conclusions from the vertex operator algebra (\ref{orderops}). First, if one demands that the string is closed we have to impose that the vertex operator is $2\pi $ periodic along $\sigma$, 
that is, $\hat{W}_{\K}(|z|,\sigma + 2\pi)= \hat{W}_{\K}(|z|,\sigma )$. 
Since $\theta$ is quasi-periodic with quasi-period $2\pi$, this condition is consistent with the vertex operator algebra if and only if  $2\eta(\lambda\K,\lambda\K')$ is integer-valued. This means that the lattice $(\Lambda,2\eta)$ of momenta $\lambda \K$, equipped with the norm $2 \eta$, must be an integer lattice.\footnote{Parameterizing as usual in terms of momentum and winding 
integers $n,w$, we get \be 
k=\frac{n}{R}, \qquad \tilde{k}=\frac{ R w}{2 \lambda^2}. \ee  Indeed, we have $2\eta(\lambda \K,\lambda \K')=nw'+wn'\in\Z$.} 
Since the staircase distribution is valued in $\pi \mathbb{Z}$ this implies that the $\sigma$-dependent phase factor is in fact independent of $\sigma$ 
\be \label{phase}
e^{-2i\eta(\lambda \K,\lambda\K')\theta(\sigma_{12})} = (-1)^{2 \eta(\lambda \K,\lambda \K')}.
\ee
Demanding that this vertex operator be $2\pi$-periodic is not enough. We also need to impose {\it mutual locality}, the condition that 
vertex operators which are space-like separated on the world-sheet commute with each other\footnote{This is usually called {\it local causality} in the Lorentzian context. Since we are dealing with the Wick rotated theory it appears as mutual locality.}. The mutual locality of vertex operators with the same momenta requires  the phase factor (\ref{phase}) to be trivial for identical momenta, that is, it requires 
 $2\eta(\lambda\K,\lambda\K)$  to be an {\it even} integer.  In other words, the lattice is such that the scalar products  $\eta(\lambda\K,\lambda\K) \in \mathbb{Z}$ are all integers, and this in turn implies\footnote{Since 
 \be
 \eta(\lambda\K,\lambda\K')  =\frac12\left( \eta(\lambda(\K+\K'),\lambda(\K+\K'))- \eta(\lambda\K,\lambda\K)- \eta(\lambda\K',\lambda\K')\right) .
 \ee} that $\eta(\lambda\K,\lambda\K') \in \mathbb{Z}/2$.  
 Therefore, the closed string boundary condition and the imposition of mutual locality of identical vertex operators  demands that the lattice $(\Lambda,2\eta)$ is an even  self-dual lattice \cite{Narain:1985jj, Narain:1986am, Polchinski:1998rq}.
 This requirement means that  we must have 
\beq\label{opeloc}
\eta(\lambda\K,\lambda\K)=2\lambda^2 (k\cdot\tk)\in\Z.
\eeq
This condition can be equivalently stated as  the demand that $(k,0)$ and $(0,\tk)$ are   elements of the lattice $\Lambda$ if $\K=(k,\tk)$ is, which means that $\Lambda $ decomposes as a direct sum of two sublattices which are Lagrangian with respect to $\eta$ and dual to each other: $\Lambda =\Gamma\oplus \tilde{\Gamma}$ with $\Gamma^*=\tilde{\Gamma}$.
 Usually, this restriction on $\Lambda$ comes from the demand of modular invariance. Here we see that it appears more naturally as the demand of mutual locality. 
Note that with our conventions (see also \cite{Polchinski:1998rq}) 
the Virasoro zero-mode generators are given by 
$
L_0 = {\alpha' (\hat{p}-\hat{\tilde{p}})^2}/{4} + N_L +1$, and $\bar{L}_0 = {\alpha' (\hat{p}+\hat{\tilde{p}})^2}/{4} + {N}_R + 1$, 
where $N_L+1$ (resp. $N_R+1$) is the number of left (resp. right) oscillators. 
The on-shell equations $(L_0,\bar{L}_0)=(1,1)$ for the string states can therefore be written in terms of the wave vectors as 
\be 
\lambda^2(k^2 + \tilde{k}^2) + N_L+{N}_R =0,\qquad 
N_L-{N}_R = 2\lambda^2 (k\cdot \tilde{k}).
\ee
The second condition, called {\it level matching}, is consistent with the condition (\ref{opeloc}) coming from mutual locality.

 We can finally impose mutual locality of  vertex operators carrying different momenta $\K$ and $\K'$. This  implies a condition on the cocycle factors that were defined in equation (\ref{defcocycle}):
\beq\label{cocyclecon}
\frac{\epsilon_{\K,\K'}}{\epsilon_{\K',\K}}=e^{2\pi i\eta(\lambda\K,\lambda\K')},
\eeq
where given the above discussion, we have $e^{2\pi i\eta(\lambda\K,\lambda\K')}=\pm 1$.
A solution to this equation is given by
\beq\label{e2}
 \epsilon_{\K,\K'}=e^{2\pi i\lambda^2\tk\cdot k'}.
\eeq
With this definition we can evaluate the LHS of (\ref{cocyclecon}), which is skew-symmetric in $(\K,\K')$ while the RHS is symmetric. The difference between the two is a factor $e^{4\pi i \lambda^2 k\cdot \tk'}$ which is equal to 1 by the condition (\ref{opeloc}). 
We note that (\ref{e2}) automatically satisfies (\ref{cocycle}), as would the exponential of any bilinear form. This is not without ambiguity, as it could be multiplied by any expression $s_{\K,\K'}\in \mathbb{Z}_2$ 
symmetric under the interchange $\K\leftrightarrow\K'$, which satisfies the cocycle condition (\ref{cocyc}). 
A change of normalization $U_\K \to \alpha_{\K} U_\K$ modifies the cocycle by the multiplication of a symmetric cocycle $\delta\alpha_{\K,\K'} =\frac{ \alpha_{\K+\K'}}{ \alpha_{\K} \alpha_{\K'}}$. 
It turns out that  any closed symmetric $\mathbb{Z}_2$  cocycle is a coboundary\footnote{{ 
This question can be phrased as the vanishing of the symmetric group cohomology 
 $H^2_S(\Lambda,\Z_2)$  \cite{Higgs}.
}}. We can therefore always make the choice (\ref{e2}) by choosing the normalization of $\hat{U}_\K$ appropriately. From now on we assume that this is done. In this case the only ambiguity left is a choice of normalization $\alpha_\K$ such that $\delta\alpha=1$, so that $\alpha_\K =  e^{i\eta(\K,\mathbb{O})}$ is a linear functional. 


Note that $\epsilon_{\K,\K'}$ can be written in a covariant form by introducing an antisymmetric form $\omega$
\beq
\omega(\K,\K'):= (k\cdot\tk'-\tk\cdot k')=\frac12(k_L\cdot k'_R-k_R\cdot k'_L),
\eeq
whereby the choice (\ref{e2}) becomes
\beq
 \epsilon_{\K,\K'}=e^{i{\pi} (\eta-\omega)(\lambda \K,\lambda \K')}.
\eeq
This is an important result. Recall that this phase appears in the product of zero mode operators
\beq\label{NCU}
\hat U_\K \hat U_{\K'}=\epsilon_{\K,\K'} \hat U_{\K+\K'},
\eeq
which corresponds to an instance of the Heisenberg group generated by $(\hat{x},\hat{\tx})$.
To proceed further, we consider representations of $\hat U_{\K}$ satisfying this algebra. 

 Up to now we have used the commutation relations of $(\hat{x},\hat{\tx})$ with $(\hat{p},\hat{\tp})$ and we have remained agnostic about the commutations of $\hat{x}$ with $\hat{\tx}$ defined as the zero mode of the string. We now want to know what commutation relations are admissible for $(\hat{x},\hat{\tx})$.
In order to analyze this issue  we introduce commuting variables  $(\hat{q},\hat{\tilde{q}})$ which stand as placeholders for $(\hat{x},\hat{\tx})$. We now  investigate whether these can be identified with the string zero modes.

To begin, suppose we take 
\beq
\hat U_\K =  e^{ik_L\cdot \hat q_L+ik_R\cdot \hat q_R} \hat{C}_{\K}
=  e^{i\tilde{k}\cdot \hat{\tilde{q}}+ik\cdot \hat{q}} \hat{C}_{\K}.
\eeq
To proceed we need to make some assumptions about the commutation relations of $\hat{q}$ and $\hat{\tilde{q}}$. In order to recover the usual framework when 
the theory is decompactified, we need to have $[\hat{q}^a,\hat{q}^b]=0$ and by duality $[\hat{\tilde{q}}_a,\hat{ \tilde{q}}_b]=0$.  Suppose that  we also make the assumption that 
\beqn
&[\hat q ,\hat{\tilde{q}} ] =0= \left[\hat q_L,\hat q_R\right].&\label{qqcomm}
\eeqn
Let us recall that we have established that $\hat{C}_{\K}$ depends only on the operators $(\hat p_{L},\hat{p}_R)$ in order to get  (\ref{reorderingJunk}), which  follows from  (\ref{HeisAlgs}). Under these assumptions, and in order to get the commutation relations (\ref{NCU}),  $\hat{C}_{\K}$ has to be an operator, 
which up to a constant factor is given by  
\beq
 \hat{C}_{\mathbb{K}} = e^{i\pi\alpha' \tk\cdot \hat{p}},
\eeq
from which  it follows that
\beq
 \hat U_{\K}= e^{ik\cdot \hat q}\, e^{i\tk\cdot(\hat{\tq}+\pi\alpha'\hat{p} )} .
\eeq
This is the result obtained, for example, in \cite{Polchinski:1998rq} and written in terms of the dual coordinates. The choice of the cocycle made in (\ref{e2}) insures that in the non-compact case, when  $\tilde{k}=0$ , no modification of the wave operator arises.  It seems very natural that the zero modes satisfy (\ref{qqcomm}), as this is in keeping with an interpretation in terms of commutative geometry, so that they (or more precisely $q=q_L+q_R$) represent coordinates on ordinary classical space-time. The downside is that cocycles appear in the algebra of physical operators. 
 So the zero mode operator algebra is {\it not} just the algebra generated 
by $q$ and $\tilde{q}$ but $q$ and $\tilde{q}+\pi\alpha' p$, a combination that depends on the choice of an O$(d,d)$ frame, as we will see in what follows.

In the following section, we will provide evidence for the following re-interpretation. We will argue that up to a normalization one should take 
\beq\label{properrep}
\hat U_\K = 
 e^{ik\cdot \hat x} e^{i\tilde{k}\cdot \hat \tx},
\eeq
where $x=x_R+x_L$ and $\tx=x_R-x_L$ are the zero mode operators appearing above in the mode expansion (\ref{mode1},\ref{mode2}).  A detailed analysis of the symplectic structure\footnote{We remind the reader that such an analysis is required to deduce Poisson brackets in the classical theory and commutation relations in the quantum theory. The result that we are claiming here is the one consistent with  locality and causality of the worldsheet theory, as we will explain in the following.} of the zero modes indicates that the operator that is canonically conjugate to $\hat\tp$ is not $\hat\tx$, but instead $\hat{\tilde{q}}:= \hat\tx-\pi\alpha' \hat p$. It is this operator, rather than $\hat{\tx}$, that commutes with $\hat{x}$. In other words, the commutative placeholders that we used in the previous section (also used in \cite{Polchinski:1998rq}) to label vertex operators are not the string zero modes. In fact, the symplectic structure implies that the zero modes satisfy
\beq\label{doublecomm}
[\hat x,\hat{\tx}]=2\pi i\lambda^2 .
\eeq
Remarkably, this commutation relation is consistent with a trivial co-cycle as in the representation (\ref{properrep}).
Also this shows that the  zero mode operator algebra is now just the algebra generated  by $\hat{x}$ and $\hat{\tilde{x}}$ with no extra input.

Thus in the case where the double coordinates satisfy (\ref{doublecomm}), the vertex operators contain no cocycle factors, but form a {\it non-commutative} subgroup of the Weyl group on the double space ${\cal P}$ coordinatized by $(x,\tx)$. This means that we have to interpret  the string zero modes as coordinates in a non-commutative  space-time. Note that eq. (\ref{doublecomm}) does not depend on the compactification radius, but only on the string length $\lambda$. Thus there is an $\alpha'$ effect present in the physics of the zero modes that is essentially implied by worldsheet locality. We will discuss  in a separate paper what might be
the fate of the usual low energy limit $\alpha'\to 0$ associated with effective field theories. 

To summarize, the zero mode operators appearing in physical vertex operators may be written as in (\ref{properrep}).
Note that in writing it this way, we have chosen an operator ordering, or equivalently we have made a choice of a pure (non-operatorial) phase. In the double space notation, $\X^A(\tau,\sigma)=(X^\mu(\tau,\sigma), \tilde{X}_\mu(\tau,\sigma))$, this can be written covariantly as 
\beq
 \hat U_{\mathbb{K}} = e^{- i\frac{\pi }{2}  \eta(\lambda \mathbb{K},\lambda \mathbb{K})} e^{i \eta(\K,\hat\X)} = e^{- i\pi  \lambda^2 (k\cdot \tilde{k})} e^{i (\tilde{k}\cdot \hat{\tx} + k\cdot \hat{x})}.
\eeq

\section{The Symplectic Structure of the Zero Modes}

In this section, we will derive the string symplectic structure. We began with the traditional presentation of the zero mode sector of the compactified string, presented either as $(x_L,p_L;x_R,p_R)$ or equivalently as $(x,p;\tx,\tp)$. Usually, these are interpreted as Darboux coordinates on a 4D-dimensional phase space, on which the symplectic form has been block off-diagonalized. That is, the zero modes form two commuting Heisenberg algebras. This interpretation would be perfectly fine if we were interested only in polynomials of these variables. However, of course we are not: we are interested in operators containing exponentials of the zero modes, as we have discussed above. Indeed, it is precisely in the usual context of compactification that we are interested in operators that are well-defined on tori. It is the algebra of these exponential operators that is important, and not the algebra of $X$s, which are not single-valued.

In this section, we will consider carefully the symplectic structure of the classical Polyakov string.\footnote{We are not aware of the foregoing analysis appearing in the existing literature. However, the result seems to have been guessed  in Ref. \cite{Sakamoto:1989ig} (and apparently not subsequently mentioned).} A systematic way to study the symplectic structure is to vary the classical action, and evaluating this on-shell, to extract the symplectic 1-form.  We will consider a cylindrical worldsheet, cut open along $\sigma=0$ to a rectangular region, coordinatized by $\s\in [0,2\pi]$, $\tau\in [\tau_0,\tau_1]$. It is important to recall that the dual field $\tilde{X}$ is defined by the identity 
$\pa_\tau X := \pa_\sigma \tilde{X}$. The dynamics of the dual field can  then simply be given by  $ \pa_\tau \tilde{X}\hat{=} \pa_\s X $, where hatted equalities mean that they are taken on-shell. 
It is obvious that these two equations imply  the usual dynamics for the string field $X$. 
 The action is
\beq\label{pol}
S=\frac{1}{4\pi \alpha'}\int d^2\sigma \left[ (\pa_\tau X)^2-(\pa_\sigma X)^2\right] .
\eeq
Here the integral $\int d^2\sigma$ is over the domain $\int_{\tau_0}^{\tau_1}\rd \tau \int_0^{2\pi} \rd \s$. 
Note that, in principle, this can be generalized to include additional terms integrated over boundaries of the worldsheet.  Varying (\ref{pol}), we find
\beqn
\delta S&=&\frac{1}{2\pi \alpha'}\int d^2\sigma\Big[ \pa_\tau (\delta X\cdot\pa_\tau X)-\pa_\sigma(\delta X\cdot\pa_\sigma X)-\delta X\cdot (\pa_\tau^2-\pa_\sigma^2)X\Big] \nonumber\\
&\hat{=}&\frac{1}{2\pi \alpha'}\int_0^{2\pi}d\s \Big(\delta X\cdot\pa_\tau X\Big)\Big|_{\tau_0}^{\tau_1}
-\frac{1}{2\pi \alpha'}\int_{\tau_0}^{\tau_1}d\tau \Big(\delta X(\tau,2\pi)-\delta X(\tau,0)\Big)\cdot\pa_\sigma {X}(\tau,0) .\nonumber
\\
&\hat{=}&\frac{1}{2\pi \alpha'}\int_0^{2\pi}d\s \Big(\delta X\cdot\pa_\tau X\Big)\Big|_{\tau_0}^{\tau_1}
-\frac{1}{2\pi \alpha'}\int_{\tau_0}^{\tau_1}d\tau \Big(\delta X(\tau,2\pi)-\delta X(\tau,0)\Big)\cdot\pa_\tau \tilde{X}(\tau,0) . \label{varS}
\eeqn
In the second line we have used the bulk equation of motion in the form $\Box X=0$ while in the third line we have used them on the boundary in the form $\pa_\tau \tilde{X}=\pa_\s X$. 
Notice that there is a term integrated along the cut in the worldsheet (which we have placed at $\sigma =0$) in addition to the usual term integrated over the space-like boundaries. This extra term appears precisely because in the present context, the fields $(X,\tilde{X})$, and their variations, are not single-valued on the worldsheet (although the current $dX$ and the dual current $d\tilde{X}$ are single-valued). 

Now, we define the charges
\beq
p_{\cal C}=\frac{1}{2\pi \alpha'}\int_{\cal C} *dX
\hat{=} \frac{1}{2\pi \alpha'}\int_{\cal C} d\tilde{X},\qquad \tp_{\cal C}=\frac{1}{2\pi \alpha'}\int_{\cal C}dX,
\eeq
where ${\cal C}$ is a  cut on the surface. The equation of motion implies that $(p_C,\tp_{\cal C}) $ depend only on the homology class of $\cal C$.  In the coordinates we are using here and for a spacelike cut,  we have
\beq
p=\frac{1}{2\pi \alpha'}\int_{0}^{2\pi}d\sigma\ \pa_\tau X,\qquad \tp=\frac{1}{2\pi \alpha'}\int_{0}^{2\pi}d\sigma\ \pa_\sigma X.
\eeq
We note that $p$ is conserved by virtue of the equation of motion and the periodicity of $\pa_\sigma X$, while $\tp$ is conserved simply because of the periodicity of $\pa_\tau X$. The last equality can be used to evaluate the last term in (\ref{varS})
\beqn
\delta \tp=\frac{1}{2\pi \alpha'}\int_{0}^{2\pi}d\sigma\ \pa_\sigma \delta X=\frac{1}{2\pi \alpha'}\Big(\delta X(\tau,2\pi)-\delta X(\tau,0)\Big).
\eeqn
This allows us to write the on-shell variation as
\beqn
\delta S &\hat{=}&\left[ \frac{1}{2\pi \alpha'}\int_0^{2\pi}d\s\ \delta X\cdot\pa_\tau X \right]_{\tau_0}^{\tau_1}
-\int_{\tau_0}^{\tau_1}d\tau\ \delta \tp\cdot\pa_\tau \tilde{X}(\tau,0)\\
&=&\left[\frac{1}{2\pi \alpha'}\int_0^{2\pi}d\s\ \delta X\cdot\pa_\tau X
- \delta \tp\cdot \tilde{X}(\tau,0)
\right]_{\tau_0}^{\tau_1} .\label{deltaSTheta}
\eeqn
We refer to the second term here as the `corner term'.
From the variation of the action, we deduce the canonical symplectic 1-form
$\Theta(\tau)$ 
defined on a spatial slice of the worldsheet at time $\tau$
\beq
\boxed{
\Theta(\tau)=- \delta \tp\cdot \tilde{X}(\tau,0)+\frac{1}{2\pi \alpha'}\int_0^{2\pi}d\s\ \delta X(\tau,\sigma)\dd 
\pa_\tau {X}(\tau,\sigma) 
\label{varPoly0}.}
\eeq

Let us remark on the presence of the corner term, which  is essential for two reasons.
First, it is responsible for ensuring diffeomorphism invariance: one can check that the corner term removes any dependence on the location of the cut. 
This can be seen by considering the symplectic potential along a cut placed at $\s_0$. This is given by 
\beq
\Theta(\tau)=- \delta \tp\cdot \tilde{X}(\tau,\s_0)+\frac{1}{2\pi \alpha'}\int_{\s_0} ^{\s_0+2\pi}d\s\ \delta X(\tau,\sigma)\dd \pa_\s \tilde{X}(\tau,\sigma),
\label{varPoly}
\eeq
where we have used the definition of $\tilde{X}$ via $\pa_\tau X =\pa_\s \tilde{X}$,  in order to write the potential in a convenient form. 
Taking the derivative of this expression with respect to $\s_0$, we can establish that it 
is indeed independent of $\s_0$. Without the corner term, the re-parametrization invariance of the worldsheet would be lost at the level of the symplectic potential. The mechanism of having in a gauge theory new degrees of freedom (edge modes) appearing at the boundaries of  regions in which the theory is defined in order to restore covariance  has been described in general terms 
in \cite{Donnelly:2016auv} and applied to gravity and Yang-Mills theory. Here we see that the same mechanism is at play, the subtlety being that even if the string is closed for the fields $\rd X$, it possesses boundaries for the field $X$ and the restoration of covariance requires the introduction of the zero mode for $\tilde{X}$.

Second,  the corner term  contribution to the symplectic structure is responsible for making $(\tx, \tp)$ dynamical. Indeed, using the  mode expansions (\ref{mode1},\ref{mode2})
we derive
\beqn
\frac{1}{2\pi \alpha'}\int_0^{2\pi}d\s\ \delta X\!\cdot\!\pa_\tau X
&=&\frac{1}{2\pi \alpha'}\int_0^{2\pi}d\s\ 
\Big(\delta x(\tau) +\alpha' \s\delta\tp  +\delta Q_+(\tau+\sigma)+\delta Q_-(\tau-\sigma)\Big)
\nonumber\\&&\qquad \qquad \quad \times
\Big(\alpha'p +\pa_\tau Q_+(\tau+\sigma)+\pa_\tau Q_-(\tau-\sigma)\Big)\\
&=&\delta x(\tau)\!\cdot\!p
+\delta\tp\cdot\!\Big(\pi \alpha' p+Q_+(\tau)-Q_-(\tau)\Big)
\\&&
+\frac{1}{2\pi \alpha'}\int_0^{2\pi}d\s\ 
\Big(\delta Q_-(\tau-\sigma)\cdot\pa_\tau Q_-(\tau-\sigma)+\delta Q_+(\tau+\sigma)\cdot\pa_\tau Q_+(\tau+\sigma)\Big)
\nonumber\\&&\label{totvar}
-\frac{1}{2\pi \alpha'}\delta\Big(\int_0^{2\pi}d\s\ Q_+(\tau+\sigma)\cdot\pa_\s Q_-(\tau-\sigma)\Big) ,
\eeqn
while the corner term in (\ref{varPoly0}) gives 
\beqn
-\delta \tp\cdot \tilde{X}(\tau,0)=-
\delta \tp\cdot\Big(\tx(\tau)+Q_+(\tau)-Q_-(\tau)\Big) .
\eeqn
Combining these two expressions we see that the zero mode factors 
contain a term $p\cdot\delta x(\tau)-\delta\tp\cdot (\tx(\tau)-\pi\alpha' p)$. We are  now ready to evaluate the symplectic potential. 
The symplectic structure is obtained by variation $\Omega=\delta \Theta$, where $\delta$ is the variational differential satisfying $\delta^2=0$. Denoting $\curlywedge$ the wedge product of variations $\delta a \curlywedge \delta b=-\delta b\curlywedge \delta a$  we get
\beqn\boxed{
\Omega(\tau)=\delta p\dot{\curlywedge}\delta x(\tau)+\delta\tp\dot{\curlywedge}(\delta\tx(\tau)-\pi\alpha' \delta p)
\label{zeromodesymp}
-\frac{1}{2\pi\alpha'}\int_0^{2\pi}d\s\ 
\Big(\delta Q_+\dot{\curlywedge} \pa_\tau\delta Q_+ + \delta Q_-\dot{\curlywedge} \pa_\tau \delta Q_-\Big).}
\eeqn
The periodic modes inside  the integral  are consistent with the commutation relations\footnote{Indeed we have that 
\be
-\frac{1}{2\pi\alpha'}\int_0^{2\pi}d\s\ 
\Big( \delta Q_- {\curlywedge} \pa_\tau \delta Q_-\Big)
= -i\hbar \sum_{m=1}^\infty\frac1m \delta \alpha_m \curlywedge \delta \alpha_{-m}.
\ee Inverting $\Omega/i\hbar$ gives the commutator $[\hat{\alpha}_m,\hat{\alpha}_{-m}]=m$.
 } for the oscillators $\alpha_m$ and $\tilde\alpha_m$. The first two terms contain the symplectic form of the zero modes. 

To summarize: the dynamics of $\tx,\tp$ comes from the corner term. Thus the corner term is responsible for the fact that $x,p,\tx,\tp$ (or equivalently $x_L,p_L,x_R,p_R$) must be thought of as independent variables. This is natural, since the physical significance of $\tx,\tp$ comes about from relaxing boundary conditions and thus should not arise from the bulk of the (constant time slice of) the worldsheet. Also, there is an additional term $\pi\alpha' \delta p\dot{\curlywedge}\delta\tp$ in the symplectic 2-form. Finally, the corner term is responsible for ensuring diffeomorphism invariance, and the corner term removes any dependence on the location of the cut.

How should we read this? The proper interpretation is that $(p,x)$ and $(\tp,\tx- \pi\alpha' p)$ are canonically conjugate pairs, and in particular, we should have  $[\hat{x},\hat{\tx}- \pi\alpha' \hat{p}]=0$. Therefore 
\be\label{NC0}
[\hat{x}^a,\hat{\tx}_b]= \pi\alpha'  [\hat{x}^a,\hat{p}_b] =2\pi i\lambda^2 \delta^a{}_b.
\ee
Thus we recover directly the results (\ref{doublecomm}) established by analyzing the structure of the vertex operator algebra.
Equivalently, we can proceed more formally, rewriting the zero mode part of the symplectic 2-form as
\beq
\Omega_0=\frac12 \Omega_{\alpha\beta}\delta\mathbb{Z}^\alpha\curlywedge\delta\mathbb{Z}^\beta ,
\eeq
where we use the ordering $\mathbb{Z}^\alpha=\{p_a,x^a,\tp^a,\tx_a\}$. The corresponding commutation relations are then of the form 
\beq 
[\hat{\mathbb{Z}}^\alpha,\hat{\mathbb{Z}}^\beta]=i\hbar \Omega^{\alpha\beta}\hat {1} ,
\eeq
where $\Omega^{\alpha\beta}$ are the components of the inverse matrix. Explicitly, we have
\beqn
\Omega=\begin{pmatrix}
0 & 1 & \pi\alpha' & 0\cr
-1 & 0 & 0 & 0\cr
-\pi\alpha' & 0 & 0 & 1\cr
0 &0 &-1 &0
\end{pmatrix}
,\qquad\Omega^{-1}=\begin{pmatrix}
0 & -1 & 0 & 0\cr
1 & 0 & 0 & \pi\alpha'\cr
0 & 0 & 0 & -1\cr
0 &-\pi\alpha' &1 &0
\end{pmatrix} ,
\eeqn
and thus
\beq
[\hat{x}^a,\hat{\tx}_b]=2\pi i\lambda^2\delta^a{}_b,\qquad
[\hat{x}^a,\hat{p}_b]=i\hbar \delta^a{}_b,\qquad
[\hat{\tx}_a,\hat{\tp}^b]=i\hbar \delta_a{}^b,\qquad
[\hat{p}_a,\hat{\tp}^b]=0
.\eeq
In particular, the operators $(\hat{x},\hat{\tx})$ appearing in the mode expansions do not commute, with the noncommutativity scale given by $2\pi\lambda^2=\pi\hbar\alpha'$. Note that $\hat p$ and $\hat{\tilde p}$ do commute, which is consistent with their diagonalization, and the labeling of states with their eigenvalues. 
Recall that there is an additional payoff if we take the non-trivial commutation relations of $\hat{x}$ and $\hat{\tx}$ into account: the mutually local vertex operators involve the simple zero mode factors
\beq
\hat{U}_\K=e^{ik\cdot \hat{x}}e^{i\tk\cdot\hat{\tx}} ,
\eeq
without additional operatorial cocycle terms. This is the  result that we suggested in the previous section. 

\subsection{Causality}

Let us examine more carefully the symplectic structure derived above. In particular, let us focus on the $p\cdot \delta\tp$ term, as its presence will no doubt be controversial. One might try to argue that in fact this term can be dropped from $\Theta(\tau)$ because if both $p$ and $\delta\tp$ are on-shell, they are time-independent. Of course, adding a time-independent quantity to $\Theta(\tau)$ does not change (\ref{deltaSTheta}). In fact, this term cannot be dropped, as it is required by causality. To see this, note that the $[\hat{x},\hat{\tx}]$ commutator appears in 
\beqn
[\hat{X}(\tau,\sigma_1),\hat{\tilde{X}}(\tau,\sigma_2)]
&=&
[\hat{x},\hat{\tx}]+\alpha' [\hat{x},\hat{p}]\sigma_2+\alpha'[\hat{\tp},\hat{\tx}]\sigma_1\\&&
+\lambda^2 \sum_{m,n\neq 0}\frac{1}{mn}e^{-i(m+n)\tau}\left[ [\hat\alpha_m,\hat\alpha_n]e^{im\sigma_1+in\sigma_2}
- [\hat{\tilde{\alpha}}_m,\hat{\tilde{\alpha}}_n]e^{-im\sigma_1-in\sigma_2}\right]\nonumber\\
&=&2\pi i\lambda^2 -2i\lambda^2\left[\sigma_{12}+\sum_{n\neq 0}\frac{e^{in\sigma_{12}}}{in}\right]\\
&=&2i\lambda^2\Big[ \pi-\theta(\sigma_{12})\Big] .
\eeqn
Recall that we are working on the interval $\sigma\in [0,2\pi]$, and $\theta(\sigma)=\pi$ for all $\sigma$ in the open interval $\sigma\in (0,2\pi)$. Thus,
the non-zero   $[\hat{x},\hat{\tx}]$ commutator gives rise to the commutativity at space-like separation of the fields $X(\tau,\sigma)$ and $\tilde X(\tau,\sigma)$, that is, it gives rise to world-sheet causality, not just for the physical vertex operators, but also for $X(\tau,\sigma)$ and $\tilde X(\tau,\sigma)$. Note that since the
 $[\hat{x},\hat{\tx}]$ commutator has led to a constant in the $[\hat{X}(\tau,\sigma_1),\hat{\tilde{X}}(\tau,\sigma_2)]$ commutator, it can be interpreted as an integration constant obtained by integrating the canonical equal-time commutator 
 \be
 [\hat{X}(\tau,\sigma_1),\pa_\tau\hat{{X}}(\tau,\sigma_2)]=[\hat{X}(\tau,\sigma_1),\pa_\s\hat{\tilde{X}}(\tau,\sigma_2)] 
 = 2\pi i\hbar \alpha' \delta(\s_{12}),
 \ee with respect to $\sigma_2$. But the only value of this integration constant consistent with causality is $\pi\alpha'\hbar$. 
   
This result was in fact found in previous work in the language of the metastring \cite{Freidel:2015pka}. There, we used the double notation  $\X^A(\tau,\sigma)=(X^\mu(\tau,\sigma), \tilde{X}_\mu(\tau,\sigma))$. Using this notation,  the equal-time commutator can be written covariantly
\beq
\left[ \hat\X^A(\tau,\sigma_1),\hat\X^B(\tau,\sigma_2)\right]
=2i
\lambda^2
\left[ \pi \omega^{AB}-\eta^{AB}\theta(\sigma_{12})\right] ,
\eeq
where $\omega$ and $\eta$ are the tensors that we used earlier in the discussion of co-cycles. In the metastring construction, $\omega_{AB}$ appeared in fact in the worldsheet action as a total derivative term. The subtlety there, complementary to the present discussion, is that the $\omega$ term had to be kept even for the closed string, because of the presence of monodromies. Here,  $\omega_{AB}$ appears as the symplectic form of the $\X$ zero modes. Our  analysis therefore confirms the results established in \cite{Freidel:2015pka} that the double space  $\cal P$ should be understood as a  {\it phase space}.

\subsection{T-duality}

Before exploring the consequences of the zero mode non-commutativity let us consider a loose end concerning the action we used so far, the usual Polyakov action (\ref{pol}).
This action is well suited for the study of strings propagating in a non-compact target. However, as we have witnessed repeatedly, when the string propagates in a compact target $\mathbb{R}^d/ \mathbb{Z}^d$, the multivaluedness of the field $X$ could  be understood as if the closed string had boundaries for the zero mode operators.
If that is the case then it should be possible to introduce ``closed string boundary terms"  that depend on the string windings. We want to understand such terms.

Another related puzzle  comes from the non-commutativity that we have just established. It is well-known that  semi-classically the non-commutative nature of two observables like $(x,p)$ normally translates into the presence of 
Aharonov-Bohm-like phases associated with topologically non-trivial
paths in phase space. It is then customary to work with a properly defined action that records naturally the presence of these phases - hence the term $\int p \rd x$ added to the Hamiltonian $\int H \rd \tau$ in the usual action that accounts for the non-commutativity of $p$ with $x$. Is it then possible to find such terms in the closed string action in order to account directly for the established non-commutativity of $(x,\tx)$?

Finally, the presence of the zero-mode $\tx$ as an edge mode of the string action suggests that we could exchange the role of $X$ and $\tilde{X}$ in the action in order to have a T-dual formulation. However, the usual form of the Polyakov string action which depends only on $X$ precludes such a formulation and  one 
usually establishes the target T-duality symmetry only at the level of the spectra of the string operators. Is it possible  to account for this directly at the level of the string action?
What is interesting is that these three loose ends can all be tied up together in one go.  
In order to see this we first focus on T-duality.

To this end, consider the form of the symplectic potential $\Theta$ as written in  (\ref{varPoly}).  We denote by $\tilde{\Theta}$ the same symplectic potential after the exchange $(X,\tilde{X})\to (\tilde{X},X)$. If T-duality was manifest we might expect that $\tilde{\Theta}= \Theta$. This is not the case and instead we obtain,  after integration by parts, the relation
%
\be\label{TtildeT}
\Theta(\tau)+\delta J(\tau) = \tilde{\Theta}(\tau)
+ 2\pi\alpha'  p\dd\delta \tilde{p},
\ee
where we have defined 
\beq
J(\tau):= -  p\cdot X(\tau,0)+\frac1{2\pi \alpha'} \int_{0}^{2\pi}d\s\ \tilde{X}(\tau,\s)\cdot \pa_\s {X}(\tau,\s).
\eeq
Let us first examine the last term in (\ref{TtildeT}). This term should not concern us since the quantization condition (\ref{opeloc}) translates into
\be
{2\pi \alpha'} p\cdot \tilde{p} \in 2\pi \hbar \mathbb{Z}. 
\ee
Thus, if we look at transformations that preserve this condition, the action is modified by a quantity which is 1 when exponentiated. This is similar to the Wess-Zumino term in the principal chiral model.
The main physical difference comes from the presence of the total variation term $\delta J$. The fact that this term is a total variation means that 
it can be reabsorbed by the addition of a boundary term to the Polyakov action  $S$. The last bit of information we need in order to identify this term comes from the fact that the functional $J$ is self-dual in the sense that if we denote by $\tilde{J}$ the functional obtained from $J$ after the T-duality transformation $(X,\tilde{X})\to (\tilde{X},X)$, we have $\tilde{J}=- J$.
Therefore,
 what we are looking for is an action $s$ such that 
$ S+ s$ has the same equation of motion as $S$ and such that 
$\delta S + \delta s = \left[\Theta +\frac12 \delta J\right]_{\tau_0}^{\tau_1}$
which is also equal to $\left[\tilde{\Theta} +\frac12 \delta \tilde{J}\right]_{\tau_0}^{\tau_1}$ modulo $2\pi\hbar \mathbb{Z}$.
Note that the addition of such a term does not modify the symplectic structure, since the latter appears as a total variation in the symplectic potential. So the conclusions about the commutation relations will not be modified.
 
Remarkably, there is a solution of this set of constraints given by 
\be\label{topol}
s(X,\tilde{X}) = \hbar \int  \omega, \qquad \omega:=\frac1{8\pi \lambda ^2 } ( \rd \tilde{X}_a \wedge \rd X^a) .
\ee 
This follows from the fact that the integrand $\omega$ is a closed form and thus
\bea
s(X,\tilde{X}) &=& \frac1{4\pi \alpha'} \int_{\tau_0}^{\tau_1}\rd \tau  \int_0^{2\pi} \rd \s \left(\pa_\s( \tilde{X} \!\cdot \!  \pa_\tau X) 
-\pa_\tau( \tilde{X}\!\cdot \!  \pa_\s X)\right)\cr
&=&  \frac1{4\pi \alpha'} \int_{\tau_0}^{\tau_1}\rd \tau (\tilde{X}(2\pi)-\tilde{X}(0)) \pa_\tau X
- \frac1{4\pi \alpha'}\left[\int_0^{2\pi} \rd \s ( \tilde{X}\!\cdot \!  \pa_\s X)\right]_{\tau_0}^{\tau_1}\cr
&=& -\frac12  \left[ J(\tau)\right]_{\tau_0}^{\tau_1}.
\eea
Therefore, we have just established that the action functional that gives the symplectic potential invariant under T-duality is the Polyakov action extended by a topological term
\be\label{Sext}
S^{\mathrm ext}_\Sigma(X,\tilde{X})= S_\Sigma(X) + \hbar \int_\Sigma \omega.   
\ee
If we modify $\tilde{X}_a $ by a  vector-valued function $\alpha_a$ which is single-valued on the worldsheet $\Sigma$, the extended action is invariant 
$S^{\mathrm ext}(X,\tilde{X}+\alpha)= S^{\mathrm ext}(X,\tilde{X})$. This means that $\tilde{X}$ is a topological degree of freedom coupled to the string. Its only dynamical components are the zero modes of $\tilde{X}$.
The presence of these topological degrees of freedom makes  the appearance of the  zero modes $(x,\tx)$ in the symplectic potential more natural, since they are already there at the level of the action.

For the non-compact string,  $X$ is single-valued on the worldsheet and $\omega$ is an exact form on $\Sigma$. Its integral vanishes and thus   $\omega$  plays no role. This is why it was never noticed before.
When the string has winding, $\int_\Sigma \omega$ does not vanish. 
In the case $\Sigma$ is a disk this integral reduces to $\frac1{2\pi\lambda^2} \int_{S^1} \tilde{X} \rd X $. In this form the topological term reduces to the usual Aharanov-Bohm phase of quantum mechanics, where $\tx/ {2\pi\lambda^2} $ plays the role of the variable conjugate to $x$, and it is the Lagrangian analog of the non-commutativity relation (\ref{NC0}).  As promised, we see that the restoration of T-duality also introduces  extra phases for paths which are non-contractible. These extra phases are necessary to account for the non-commutativity of the zero modes, and these extra phases enter the action via a topological coupling that involves a symplectic structure $\omega$  in the double  space $(x,\tilde{x})$, making it into a phase space.
As we already mentioned the same conclusion was reached in the metastring formulation \cite{Freidel:2015pka}.

\subsection{$O(d,d)$ and more General Backgrounds}\label{sec:odd}

So far our discussion has been centered on the flat closed string with trivial metric and zero $B$ fields and we now want to turn on constant $G,B$ background fields
\be
S_{G,B} =\frac{1}{4\pi \alpha'} 
\int \rd^2\s\left( G_{ab}[(\pa_\tau X^a \pa_\tau X^b - \pa_\s X^a\pa_\s X^b] 
+  2 B_{ab} \pa_\tau X^a \pa_\s X^b  \right) .
\ee
As we are going to see the main effect 
that comes from the presence of a B-field is to produce a non-commutativity of the dual coordinates. 

 We define the dual coordinate $\tilde{X}_a$ as a covector where $\pa_\s \tilde{X}_a := G_{ab}\pa_\tau X^a$, while the string equation of motion  implies $\pa_\tau \tilde{X}_a \hat{=} G_{ab}\pa_\s X^b$ when $G,B$ are constant. Repeating the calculation with constant $G,B$ background fields included, one finds
\beqn
\delta S
&\hat{=}&\frac{1}{2\pi\alpha'}\int_0^{2\pi}d\sigma\ \delta X^a \left.\left(\pa_\s \tilde{X}_a+B_{ab}\pa_\sigma X^b\right)\right|_{\tau_0}^{\tau_1}-\int_{\tau_0}^{\tau_1}d\tau\ \tilde{p}^a \left(\pa_\tau \tilde X_a+B_{ab}\pa_\tau X^b\right).
\eeqn
Here we have introduced, as before, momenta $p_a= \frac{1}{2\pi\alpha'}\int_0^{2\pi}d\sigma\ \pa_\s \tilde{X}_a$ and dual momenta $\tilde{p}^a= \frac{1}{2\pi\alpha'}\int_0^{2\pi}d\sigma\ \pa_\sigma X^a$. 
We then deduce
\beqn
\Theta= \delta x^a p_a-\delta\tp^a \left[ \tx_a-\pi\alpha' (p_a+B_{ab}\tp^b)\right]+\frac{1}{2\pi\alpha'}\int_0^{2\pi}d\sigma\ \left( \delta Q^a_-(G-B)_{ab}\pa_\tau Q^b_-+\delta Q^a_+ (G+B)_{ab}\pa_\tau Q^b_+\right).\nonumber
\eeqn

We should note that the momentum field conjugate to $X$ is now
$
2\pi\alpha' P=G\cdot\pa_\tau X+B\cdot\pa_\sigma X.
$
So in this notation, $p$ is not the zero mode of $P$. Instead, we have 
\beq
\frac{1}{2\pi}\int_0^{2\pi}d\sigma\ P_a=
 p_a+B_{ab} \tp^b ,
\eeq
and it is this zero mode, i.e., the  momentum conjugate to $x$, that appears in the shift of $\tx$ in the  zero mode contribution to $\Theta$. 
The zero mode part of the symplectic form is
\beqn
\Omega&=&  \delta p_a \curlywedge \delta x^a
+ \delta\tp^a \curlywedge  \left[ \delta \tx_a-\pi\alpha' (\delta p_a+B_{ab}\delta \tp^b)\right].
\eeqn
 This symplectic one-form can be inverted in order to obtain  the commutation relations
$[x^a,p_b]=i\hbar= [\tx^a,\tp_b]$, while the commutation relations for the zero modes $(x,\tx)$ are given by\footnote{This algebra is well-known in the open string sector in the limit of non-commutative field theory \cite{Douglas:1997fm, Connes:1997cr, Seiberg:1999vs, Douglas:2001ba, Szabo:2001kg, Grosse:2004yu}.} 
\beq\label{Bcommrel}
[\hat{x}^a,\hat{x}^b]=0,\qquad 
[\hat{x}^a,\hat{\tx}_b]=2\pi i\lambda^2 \delta^a{}_b,\qquad 
[\hat{\tx}_a,\hat{\tx}_b]=-4\pi i\lambda^2 B_{ab}.
\eeq
The key property satisfied by this bracket  is that the combinations 
$\hat{x}^a$ and $\hat{\tx}_a  - \pi\alpha'(\hat{p}_a+B_{ab}\hat\tp^b) $
generate a  {\it commutative} sub-algebra.
%
The presence of a non-trivial $B$-field implies that the subalgebra generated by the dual coordinates $\tx$ is non-commutative, with non-commutativity being proportional to the  background $B$-field. 

In order to understand the nature of this result it is convenient to introduce the following {\it kinematical} tensors on doubled space
\beq\label{etaom}
\eta_{AB}=\begin{pmatrix}0&\delta_a{}^b\cr \delta^a{}_b&0\end{pmatrix}
,\qquad
\omega_{AB}=\begin{pmatrix}0&-\delta_a{}^b\cr \delta^a{}_b&0\end{pmatrix},
\eeq
where $\eta$ defines a neutral $O(d,d)$ metric and $\omega$ a compatible 2-form. The compatibility means that the tensor $K:=\eta^{-1}\omega$ is a product structure, i.e., it squares to the identity $K^2=1$.
When we set  $B_{ab}=0$, then we can simply write the zero mode symplectic
form  in a covariant manner as 
\beq
\Omega =  \eta_{AB} \delta{\mathbb P}^A\curlywedge\delta \X^B+\frac{\pi\alpha'}2  \omega_{AB} \delta {\mathbb P}^A\curlywedge \delta{\mathbb P}^B.
\eeq
When we turn on  the $B_{ab}$ field, there are two ways to write the symplectic form in covariant form. The first one, which we call the {\it passive} point of view,  is to consider that a non-zero $B$-field corresponds to a change of {\it background}.

 One  introduces the modified metric and form
\beq
\eta^{(B)}_{AB}=\eta_{AB}=\begin{pmatrix}0&\delta_a{}^b\cr \delta^a{}_b&0\end{pmatrix},\qquad
\omega^{(B)}_{AB}=\begin{pmatrix}-2B_{ab}&-\delta_a{}^b\cr \delta^a{}_b&0\end{pmatrix} ,
\eeq
and using these we can simply evaluate the symplectic structure in a $B$ background as 
\beqn\label{OmwithB}
\Omega= \eta_{AB}  \delta {\mathbb P}^A\curlywedge \delta \tilde\X^B+\frac{\pi\alpha'}2 \omega^{(B)}_{AB} \delta{\mathbb P}^A \curlywedge \delta{\mathbb P}^B\label{Thetapassive}.
\eeqn
In this passive picture the change of background results in a modification of the commutation relations of $\X$, which can be written as 
\beq
\left[\hat\X^A,\hat\X^B\right]=2\pi i\lambda^2 (\omega^{(B)})^{AB},\quad\mathrm{where}\quad
(\omega^{(B)})^{AB}=\begin{pmatrix}0&\delta^a{}_b\cr -\delta_a{}^b&-2B_{ab}\end{pmatrix}.
\eeq
This is equivalent to  (\ref{Bcommrel}).
For this reason, we see that $\omega_{AB}$ plays the role of a symplectic form on the subspace of phase space coordinatized by $(x,\tx)$, and we will refer to it as such. 

Thus we realize that the modified forms are simply given by a change of the O$(d,d)$ frame 
\beq
\eta^{(B)}=(e^{\hat{B}})^T \eta e^{\hat{B}} =\eta ,\qquad
\omega^{(B)}=(e^{\hat{B}})^T \omega e^{\hat{B}} ,
\eeq
where $e^{\hat{B}}$ is the $O(d,d)$ transformation determined by the $B$-field.\footnote{The transformation $\hat{B}(x,\tilde{x}) = (0,B(x))$ is a nilpotent transformation and its exponential is linear in $B$.}

The alternative {\it active} interpretation of (\ref{OmwithB}) is to keep the background fields unchanged and to modify the  $O(d,d)$ frame by an active transformation $e^{\hat{B}}$, i.e., one that acts on $\Pm$ and $\X$ as 
\beq
\X^A=
\begin{pmatrix}x^a\cr \tx_a\end{pmatrix}\to 
(e^{\hat{B}}\X)^A
=\begin{pmatrix}x^a\cr \tx_a+B_{ab}x^b\end{pmatrix},
\qquad
{\mathbb P}^A=\begin{pmatrix}\tp^a\cr p_a\end{pmatrix}
\to (e^{\hat{B}}{\mathbb P})^A 
=\begin{pmatrix}\tp^a\cr p_a+B_{ab}\tp^b\end{pmatrix},
\label{XtransOdd}
\eeq
which results in the symplectic form
\beq \label{passive}
\Omega=\eta(e^{\hat{B}}\delta\Pm,e^{\hat{B}}\delta \X)-\frac{\pi\alpha'}2 \omega(e^{\hat{B}}\delta \Pm, e^{\hat{B}}\delta\Pm).
\eeq
Of course, the active and passive transformations have the same effect on $\Omega$, as (\ref{OmwithB}) is equal to (\ref{passive}). We emphasize that although the idea that a $B$-field yields an $O(d,d)$ transformation is familiar, here we see that because we have appreciated the significance of the form $\omega$, the $O(d,d)$ transformation is entirely geometric.

A similar perspective is found in the dynamical sector where the spectrum is given by 
\be
{\cal H}=\lambda^2 H_{AB} \hat{\mathbb{K}}^A\hat{\mathbb{K}}^B + N_L + \tilde{N}_R -2, \quad\mathrm{with}\quad H_{AB}= \begin{pmatrix}G_{ab}&0\cr 0&G^{ab}\end{pmatrix},
\ee
where $\hat\Pm =\hbar \hat{\mathbb{K}}$ and $N_L,{N}_R\in \mathbb{N}$ are the oscillator numbers.
This spectrum is  modified in the presence of the $B$-field by a redefinition 
$H \mapsto H^{(B)}$ (or equivalently, by a change of frame $\Pm \to e^{\hat{B}} \Pm$). 
The new ingredient here is that this field redefinition also affects the kinematics of the effective string geometry by changing the nature of the non-commutativity of the zero modes.

That the introduction of a $B$-field implies a modification of the canonical structure of zero modes can also be understood from the point of view of the extended action (\ref{Sext}). 
Under a $B$ transform the topological action term is modified to
\be
\hbar \int \omega^{(B)} = \frac{1}{4\pi\alpha'}
\rd X \wedge \rd (\tilde{X}_a + B_{ab}X^b)
= \hbar \int \omega + \frac{1}{4\pi\alpha'}
\int  B_{ab} \rd X^a \wedge \rd X^b .
\ee
Therefore, we see that we can write the extended action either as 
$ S^{\rm ext}_{G,B} = \hbar \int \omega +  S_{G,B}$ or as 
\be
 S^{\rm ext}_{G,B} = \hbar \int \omega^{(B)} +  S_{G,0}.
\ee
In the second formulation, which corresponds to a different choice of the O$(d,d)$ frame, we see that the effect of the B-field is simply to modify the topological interaction term. 
The fact that the geometrical elements of the string given by the pair $(\eta,\omega)$, which controls the kinematics of the theory, and $H$, which controls the metric, are all rotated in a different frame into $(\eta, \omega^{(B)}, H^{(B)})$, implies that $(\omega, \eta, H)$ form the elements of a Born geometry \cite{Freidel:2014qna}.

More generally, let us recall that a general O$(d,d)$  transformation 
$g: (x,\tx) \to (A x+ C\tx, D\tx + Bx)$, such that $A$ is invertible,
can be decomposed uniquely as a product $g = e^{\hat{B}} \hat{A} e^{\hat{\beta}}$ of a 
 B-transformation $e^{\hat{B}}(x,\tilde{x}) = (x,\tx + B(x))$ labeled by a 2-form, a GL$(d)$ transformation $\hat{A} (x,\tx) = (A x, (A^T)^{-1} \tx)$, where $A \in \mathrm{GL}(d)$, and a $\beta$-transform 
 $e^{\hat\beta}(x,\tilde{x}) = (x+\beta(\tilde{x}),\tx )$ labeled by a skew symmetric bi-vector $\beta^{ab}$. These sets of transformations do not include the T-duality transformations $T_a$ which exchange $h_{ab}x^b $ and $\tx_a$ (with $h_{ab}$ the Minkowski metric). 
The $A$-transform and the $B$-transform are special in the sense that they do not affect our notion of  space-time. The usual space-time can be embedded in the double space as the slice $M:=\{(x,0)\}$. Both $A$- and $B$-transformations map the set $M$ onto itself and this explains why these have a commutative space-time interpretation.
On the other hand, the $\beta$-transform and T-duality do not preserve the space-time slice $M$, mixing $x$ with 
$\tilde{x}$. Since these coordinates do not commute, the proper way to understand these transformations is in the context of non-commutative geometry.

%

\section{Conclusion}

In this paper we have uncovered an intrinsic non-commutative structure in closed  string theory on flat backgrounds.
For a compactified string, we have found that the doubled space of zero modes coordinatized by $(x,\tx)$ is in fact a non-commutative geometry. The precise form of this non-commutativity is that it is a Heisenberg algebra. From this point of view, the zero mode space can be considered as a phase space. Drawing an analogy with \cite{Freidel:2016pls} then, it seems natural to conjecture that in this framework a choice of space-time corresponds to a choice of commutative subalgebra. The obvious classical choices of commutative subalgebras corresponding to the subspace coordinatized by $\{x^\mu\}$ (or $\{\tx_\mu\}$), as would be recovered in decompactification limits. On the other hand, in Ref. \cite{Freidel:2016pls} we pointed out the existence of {\it quantum polarizations} of the Heisenberg group given by a choice of modular variables. In this case, the length scale inherent to such a modular polarization is set by the string scale $\lambda$ itself. 

From the string point of view, space-time, which is the target space of string theory, is the usual commutative compactified space. However, from the effective field theory point of view, space-time is 
an index space. Our point is twofold. First, space-time viewed as an index space does not have to be the same as the space-time viewed as a target. This has already been observed in the context of double field theory \cite{Hull:2009mi}. The second point is that, in order to understand space-time as an index space, we have to leave the realm of commutative geometry and enter the realm of non-commutative geometry. In this context, the space-time viewed as an index space appears as a commutative subalgebra of the Heisenberg algebra. From this viewpoint, strings propagating on a compact space-time involve a choice of (modular) polarization for an effective description based on string zero modes. 
Generally, T-dualities act within this framework as the operations that change the choice of polarization. 

It is natural to ask in what sense effective field theories can be thought to emerge from string theory in the context of compactification. We will explore this question elsewhere, but it seems natural to note that such field theories may be defined not in the UV, but in terms of a notion of self-dual fixed points.
Such self-dual fixed points occur for example in the doubled renormalization group \cite{Grosse:2004yu}, found in the toy models of non-commutative field theory \cite{Douglas:1997fm, Connes:1997cr, Seiberg:1999vs, Douglas:2001ba, Szabo:2001kg, Grosse:2004yu}.

Clearly, the non-commutativity that we have uncovered here will be of crucial importance for duality-symmetric formulations of string theory. In particular, as we will also discuss elsewhere, the methods that we have developed here will serve to make the structure of exotic backgrounds (such as asymmetric orbifolds \cite{Narain:1986qm} or more generally T-folds \cite{Hull:2006va}) transparent.


We conclude with a comment about the literature: there exist old papers that are apparently not widely known
in the community \cite{Sakamoto:1989ig, Erler:1991an, Sakamoto:1992ur, Horiguchi:1992sn, Sakamoto:1993bc, Sakamoto:1994nx}, in which the central results concerning the co-cycles of closed string vertex operators (in orbifold backgrounds) that follow from the presentation of this paper were also discussed. The relevance of co-cycles was more recently emphasized in \cite{Landi:1998ii, Hellerman:2006tx}. Non-commutative aspects of the vertex operator algebra of free string theory in
the zero mode sector were also studied in \cite{Frohlich:1993es}.
What was missing in these papers was the recognition of the generic non-commutativity of closed string theory as
well as the relevance of the more general structures associated with Born geometry.

\bigskip

{\bf Acknowledgements:}
{\small RGL} and {\small DM} thank Perimeter Institute for
hospitality. {\small LF}, {\small RGL} and {\small DM} thank the Banff Center for
providing an inspiring environment for work and the Julian Schwinger Foundation for support. {\small RGL} is supported in part by
the U.S. Department of Energy contract DE-SC0015655 and
{\small DM}
by the U.S. Department of Energy
under contract DE-FG02-13ER41917. Research at Perimeter Institute for
Theoretical Physics is supported in part by the Government of Canada through NSERC and by the Province of Ontario through
MRI.


\providecommand{\href}[2]{#2}\begingroup\raggedright\endgroup

\end{document}